\documentclass[12pt]{article}
\usepackage{amsmath}
\usepackage{amssymb}
\usepackage{amsthm}
\usepackage[titletoc,title]{appendix}
\usepackage{graphicx}
\usepackage{float}
\usepackage{wrapfig}
\numberwithin{equation}{section}
\usepackage[left=0.95in,right=0.95in,bottom=1.45in]{geometry}
\usepackage{setspace}
\usepackage [linktocpage]{hyperref}
\begin{document}
\title{\bf Solutions For Scalar Equations in AdS$_4$ with Adomian Method and Boundary CFT$_3$ Duals \ \\ \ }
\author{{\bf M. Naghdi\,$^{a, b}\,$\footnote{E-Mail: m.naghdi@ilam.ac.ir} } \\
$^a$\,\textit{Department of Physics, Faculty of Basic Sciences}, \\
\textit{University of Ilam, Ilam, Iran}  \\
$^b$\,\textit{School of Particles and Accelerators}, \\
\textit{Institute for Research in Fundamental Sciences (IPM)},\\
\textit{P.O.Box 19395-5531, Tehran, Iran}}
\date{\today}
 \setlength{\topmargin}{0.0in}
 \setlength{\textheight}{9.2in}
  \maketitle
  \vspace{-0.1in}
    \thispagestyle{empty}
 \begin{center}
   \textbf{Abstract}
 \end{center}
For a nonlinear partial differential equation for (pseudo)scalars in the bulk of Euclidean $AdS_4$, arising from a truncation of 11-dimensional supergravity over $AdS_4 \times S^7/Z_k$, we use math tools and in particular Adomian Decomposition Method, with initial data from near the boundary behavior of a special or general solution, although we focus on normalizable modes and Dirichlet boundary condition, to get perturbative series solutions (of the equation valid in probe approximation) for three special modes of $m^2=4, 0, -9/4$. Meantime, we remind that for the skew-whiffed M2-branes background, there are Higgs-like (pseudo)scalars that make the equation homogeneous and provide spontaneous symmetry breaking. Then, with the setups and solutions in the bulk, where all supersymmetries and parity are broken, we swap the three fundamental representations of $SO(8)$ for gravitino, deform the ABJM-like three-dimensional boundary actions with various corresponding $SU(4) \times U(1)$-singlet operators made of fermions, scalars and $SU(N)$ gauge fields, find new $SO(4)$-invariant instantons, and finally adjust the bulk and boundary solutions and confirm state-operator AdS$_4$/CFT$_3$ correspondence.

\newpage
\setlength{\topmargin}{-0.7in}
\pagenumbering{arabic} 
\setcounter{page}{2} 


\section{Introduction}
From truncations of 11-dimensional (11D) supergravity (SUGRA), with (anti)membranes wrapping around mainly the internal directions of $CP^3 \ltimes S^1/Z_k$ in (anti)M2-branes background with 4-form fluxes, we have already gotten localized objects and mainly instantons in the bulk of 4D Euclidean anti-de Sitter space ($EAdS_4$); see for instance \cite{Me5}, \cite{Me6}, \cite{Me7}. However, it is not always possible to find closed solutions for the resulting equations, which are often Nonlinear Partial Differential Equations (NPDEs) for (pseudo)scalars in the bulk; although when taking backreaction, we already found \cite{Me8}, \cite{Me9} instantons form solving exactly the equations resulting from zeroing the energy-momentum tensors of the Einstein's equations (as topological objects should not change the background geometry) together with the main equation in the bulk.

Here we ignore the backreaction, that is working in probe approximation, and try to solve the main bulk equation for three special modes of $m^2=+4, 0, -9/4$ in $AdS_4$ space. In particular, we find out that in skew-whiffed (SW) M2-branes background, the Higgs-like (massive: $m^2\geq 0$) (pseudo)scalars provide spontaneous symmetry breaking (SSB) (inherent in our truncation) and the equation becomes homogeneous; and that for the ABJM \cite{ABJM} SW background, the $m^2=+4$ mode emerges.	

Then, after we present a few general perturbative solutions-useful also for the boundary field theory analyzes-for the equation after SSB, in particular we apply the Adomian Decomposition Method (ADM) to find solutions for the NLPDEs in $EAdS_4$. Indeed, with the latter method, using the initial data from the (pseudo)scalar behavior near the boundary for the three boundary conditions (BCs) of Neumann, Dirichlet and mixed (although here we concentrate on normalizable modes and Dirichlet BC), we get series solutions near the boundary that, under AdS$_4$/CFT$_3$ correspondence rules, adjust with the boundary solutions in 3D Chern-Simon (CS) matter theories of the ABJM type.

On the other hand, with respect to (wrt) the symmetries of the bulk setups and solutions, which point out that dual boundary operators should be $SU(4) \times U(1)$-singlet, and all supersymmetries, parity- and scale-invariance must be broken, we swap or interchange the three representations (reps) of $SO(8)$ for gravitino so that, we present the needed singlet boundary scalar, fermion, (mainly $U(1)$ and $SU(2)$) gauge fields and especially the operators made of them; and then, by deforming the boundary actions with the new built operators corresponding to the three bulk states, while proposing new solutions for the boundary scalar, fermion and gauge fields, we find various interesting $SO(4)$ invariant solutions, which in turn are often small instantons on a three-sphere with radius $r$ at infinity ($S_\infty^3$), and confirm the state-operator correspondence for all the bulk and boundary solutions.	

The structure of this article is as follows: In section 2, we present the 7-form ansatz of the gravity background and main resulting nonhomogeneous NLPDE for (pseudo)scalars in $EAdS_4$; and in subsection 2.1, we see that in anti-M2-branes background, the (pseudo)scalars become Higgs-like and leave the equation homogeneous. In section 3, we try to solve the bulk equations, in probe approximation; and while in subsection 3.1 we write solutions for the equation after SSB, in subsection 3.2 we present a general procedure to solve the bulk equations with ADM, perturbatively; Then, in subsections 3.3-5, we get series (normalizable) solutions for the bulk massive ($m^2=+4$), massless ($m^2=0$) and tachyonic ($m^2=-9/4$) modes with ADM, respectively. In section 4, we review dual symmetries and needed elements of AdS$_4$/CFT$_3$ correspondence for the (pseudo)scalars, briefly. In section 5, we deal with the boundary dual Euclidean solutions, realized in ABJM-like CFT$_3$'s; and in this way, in subsections 5.1-3, after presenting dual $\Delta_+=4, 3, 3/2$ operators and deforming the associated actions, we write boundary solutions and confirm the gravity/gauge duality for all three bulk modes, respectively. In Section 6, after giving a summary and conclusion, we comment briefly on connections of these studies to hadrons and specially meson spectra in context of holographic QCD, as an important application of gauge/gravity duality.

\section{The Gravity Background and Equation}\label{section02}
We use 11D SUGRA over $AdS_4 \times CP^3 \ltimes S^1/Z_k$ with the ansatz
\begin{equation}\label{eq01}
  G_7 = R^7\, f_1\, \mathcal{E}_4 \wedge J \wedge e_7 + R^5\, \ast_4 df_2 \wedge J^2 +  R^7\, {f}_3\, J^3 \wedge e_7,
\end{equation}
where $R=2 R_{AdS}$ is the $AdS$ radius of curvature, $\mathcal{E}_4$ is for the unit-volume form on $AdS_4$, $J$ is the K\"{a}hler form on $CP^3$, $e_7$ is the seventh vielbein and $f_1, f_2, f_3$ are scalar functions of the bulk coordinates.

From the Euclidean 11D equation $d G_7 = ({i}/{2})\, G_4 \wedge G_4$, with $G_4=\ast_{11} G_7$, we get
\begin{equation}\label{eq02}
     \Box_4 f_2 - m^2\, f_2 \pm \delta\, f_2^2 - \lambda\, f_2^3 \pm F=0,
\end{equation}
see also \cite{Me9}, where $\Box_4$ is the $EAdS_4$ Laplacian,
\begin{equation}\label{eq02a}
m^2=\frac{4}{R^2} \left(1 \pm 3\, C_1 + 288\, C_2^2 \right), \quad \delta = \frac{288}{R}\, C_2,  \quad \lambda=24, \quad F = \frac{16}{R^3}\left(C_2 \pm 3\, C_2\, C_1 + 96\, C_2^3 \right),
\end{equation}
$C_1, C_2$ are real constants and
\begin{equation}\label{eq03}
 f_3 = i\, 32 R\, f_1^2\, \pm\, i\, \frac{C_1}{R},
\end{equation}
where the upper and lower signs, behind the terms including $C_1$, point out considering the Wick-Rotated (WR) and SW backgrounds respectively \footnote{We remind that the WR and SW backgrounds of the ABJM model \cite{ABJM}, with the 4-form flux of $G_4^{(0)}= \pm i N \mathcal{E}_4$, $N=(3/8) R^3$ are realized with $C_1=1$, respectively.}; and that we have
\begin{equation}\label{eq04}
f_1 = - \frac{1}{4} f_2 \pm \frac{C_2}{R},
\end{equation}
from the Bianchi identity ($dG_4=0$), where in turn the upper sign, behind the terms including $C_2$, shows that the true vacuum sits on the right-hand side (RHS) of the false vacuum, in the corresponding double-well potential, and conversely for the lower sign. Now, we see from (\ref{eq02a}) that for various values of $C_1$ and $C_2$, a tower of tachyonic (just for the SW version), massless and massive (pseudo)scalars are accessible.

In this way, while in \cite{Me9} we mainly studied the conformally coupled (CC) (pseudo)scalar $m^2 R_{AdS}^2=-2$ when including the backreaction, and the massive mode of $m^2 R_{AdS}^2=+4$ because of its relevance to the former when discussing the boundary dual solutions, however, here we focus on the bulk equation of (\ref{eq02}), valid in probe approximation, and try to solve it for the latter mode besides the massless $m^2 R_{AdS}^2=0$ one and in particular $m^2 R_{AdS}^2=-9/4$ as the lowest physically allowable mode, and study their boundary duals.

\subsection{A Special Case: Higgs-Like (pseudo)Scalars} \label{sub02-01}
From (\ref{eq02a}) we can write (with $R_{AdS}=1$)
\begin{equation}\label{eq07}
 F=\frac{\delta}{72} \left(m^2 -\frac{\delta^2}{108} \right),
\end{equation}
and so, to make the equation (\ref{eq02}) homogeneous ($F=0$), we have to set
\begin{equation}\label{eq08a}
C_2^2 = \frac{\mp 3\, C_1 -1}{96}.
\end{equation}
As a result, the mass of (\ref{eq02a}) reads
\begin{equation}\label{eq08b}
 m^2 = -2\, (1 \pm 3\, C_1) \equiv -2\, \bar{m}^2,
\end{equation}
where $\bar{m}^2$ is indeed the mass-squared for $f_1$, whose associated equation includes the coupling of $\bar{\lambda}=384$ instead of $\lambda=24$ for $f_2$. Therefore, the scalar equation (\ref{eq02}) (with $f_2=f$ from now on) reduces to
\begin{equation}\label{eq09}
 \Box_4\, f - m^2\, f \pm 6\, \sqrt{3}\, m\, f^2 - \lambda\, f^3 = 0.
\end{equation}
Next, note that although it is possible to achieve tachyonic modes for $m^2$ in the WR version of (\ref{eq02}), but it causes an imaginary $\delta$ coefficient that is not physically permissible and so, just for the SW version and with $C_1\geq 1/3$, we have acceptable modes in the case; and also that for the exact ABJM background ($C_1=1$), we have $m^2 = +4$ ($C_2 =1/\sqrt{48}$).

It should also be noted that the (pseudo)scalar here is \emph{Higgs-like} in that having (\ref{eq04}), $\pm (C_2/2)$'s are indeed for the homogeneous vacua with $C_2/2 = \upsilon = \sqrt{{- \bar{m}^2}/\bar{\lambda}}$; and that this \emph{spontaneous symmetry breaking} caused by (\ref{eq04}), in which $f$ acts as fluctuation around the homogeneous vacua, is not imposed by hand but it come from our original ansatz and reduction, automatically.

\section{Solving the Bulk Scalar Equation} \label{sub03-01}
Although a nontrivial closed solution for the NLPDE of (\ref{eq02}) in $EAdS_4$ is not accessible, however, in this section, we look for approximate or perturbative mathematical solutions suitable for boundary considerations, mainly with ADM \cite{Adomian1994}. In this way, besides taking the equation (\ref{eq09}) for the massive modes and especially $m^2=+4$ in the SW background, we consider the equations of the massless $m^2=0$ and tachyonic $m^2=-9/4$ modes and present interesting solutions.

To continue, we write the main equation (\ref{eq02}), with $f=(u/R_{AdS})\, g$ and $R_{AdS}=1$, as \footnote{From now on, we use the upper sign for the $f^2$ term in the equations.}
\begin{equation}\label{eq25bb}
     \left(\partial_i \partial_i + \partial_u \partial_u \right)\, g -\frac{(2+m^2)}{u^2}\, g + \frac{\delta}{u}\, g^2- 24\, g^3 + \frac{\delta}{216\, u^3}\, (2\, \bar{m}^2+m^2)=0,
\end{equation}
where
\begin{equation}\label{eq25ab}
     \delta = 12\, \sqrt{\frac{m^2-\bar{m}^2}{2}}, \quad \Box_4 f = \frac{u^2}{R_{AdS}^2} \left(\partial_i \partial_i + \partial_u \partial_u - \frac{2}{u} \partial_u \right) f,
\end{equation}
and we use the upper-half Poincar$\acute{e}$ metric of
\begin{equation}\label{eq27}
 ds^2_{EAdS_4} = \frac{R_{AdS}^2}{u^2} \left(du^2 + dx^2 + dy^2 + dz^2 \right).
\end{equation}

\subsection{Solutions For the Massive Equation}
We can rewrite the equation (\ref{eq09}), for Higgs-like (pseudo)scalars, as
\begin{equation}\label{eq26}
     \left[ \left(\frac{\partial^2}{\partial r^2} + \frac{2}{r} \frac{\partial}{\partial r} \right) + \left(\frac{\partial^2}{\partial u^2} - \frac{2}{u} \frac{\partial}{\partial u} \right) - \frac{m^2}{u^2} \right] f(u,r) + \frac{1}{u^2} \left[6 \sqrt{3}\, m\, f(u,r)^2 - 24\, f(u,r)^3 \right]=0,
\end{equation}
where $r= |\vec{u}|$, $\vec{u}=(x,y,z)$, and we discard the angular parts of the 3D spherical Laplacian.

A general solution for the linear part of the equation (\ref{eq26}) (say $f_0$) is in Hyperbolic (for the $r$ part)- and Bessel (for the $u$ part)- functions of the first and second kinds. Next, if we take the leading solution $f_0(u,r)$ to gain a first-order one (say $f_1(u,r)$), we can get the simplified solution of \footnote{Note that we in general take
\begin{equation}\label{eq28a}
f^{(n)} = \sum_{n=0}^n f_n.
\end{equation}}
\begin{equation}\label{eq28}
f^{(1)}(u,r) = \frac{{u}^{3/2}}{r} \left[\tilde{C}_1\,\sinh(r) +\tilde{C}_2\,\cosh(r) \right] \left[ \tilde{C}_3\, J_\nu (u) + \tilde{C}_4\, Y_\nu (u) \right],
\end{equation}
where $\sqrt{9 + 4\, m^2}=2 \nu$, $\tilde{C}_1, \tilde{C}_2, ...$ are constants, and $J_\nu (u)$ and $Y_\nu (u)$ are Bessel functions of the first and second kind respectively, with near the boundary ($u = 0$) behavior of
\begin{equation}\label{eq28aa}
J_\nu(u\rightarrow 0) \approx  \frac{1}{\Gamma(\nu+1)} \left(\frac{u}{2} \right)^\nu, \quad Y_0 (u \rightarrow 0) \approx \frac{2}{\pi} \left( \ln(\frac{2}{u}) + \gamma \right), \quad Y_\nu (u \rightarrow 0) \approx \frac{\Gamma(\nu)}{2} \left(\frac{2}{u} \right)^\nu,
\end{equation}
for $\nu$ as a nonnegative integer, and $\gamma \cong 0.577$ is the Euler–Mascheroni constant; And as a result, from (\ref{eq28}) we read the right behavior one expects for (pseudo)scalars near the boundary, that is
\begin{equation}\label{eq50ee}
f(u\rightarrow 0, \vec{u}) \approx \alpha(\vec{u})\, u^{\Delta_-} + \beta(\vec{u})\, u^{\Delta_+},
\end{equation}
where $\Delta_-$ and $\Delta_+$ are the smaller and larger roots of $m^2=\Delta (\Delta-3)$ in $AdS_4$, respectively. \footnote{It is notable that with $\tilde{C}_1 = -\tilde{C}_2$, the $r$ part solution of (\ref{eq28}) behaves as $\sim \exp(-r)/r$, which goes to zero as $r \rightarrow \infty$; It is interesting in that there is the same exponential behavior for the constrained instantons of the so-called $\phi^4$ model; see for instance \cite{KubyshinTinyakov01} and \cite{Me6} and also \cite{Me9}, where we discussed a similar (free massive) solution about critical boson model in three boundary dimensions.}

In particular, we can find an interesting solution from studying symmetries of the equation, the so-called \emph{Lie-group approach}; see, for instance, \cite{Polyanin2012}. For the equation of (\ref{eq26}), the group generator is $X= u\, \partial_u + r\, \partial_r$, which writes to an uniform extension of the coordinates, and the invariants are $I_1= u\, r^{-1}$, $I_2=f$. Therefore, by taking $I_2=\mathcal{F}(I_1)$ that means
\begin{equation}\label{eq39a}
    f(u,r)=\mathcal{F}(\xi), \quad \xi = \frac{u}{r},
\end{equation}
and indeed gives a self-similar solution, we have
\begin{equation}\label{eq39b}
 \left[ \left({\xi}^{2}+1 \right) {\frac{{d }^{2}}{{d}{\xi}^{2}}} - \frac{2}{\xi}\,\frac{d}{d \xi} - \frac{m^2}{\xi^2} \right] \mathcal{F}(\xi) + \frac{1}{\xi^2} \left[ 6 \sqrt{3}\, m\, \mathcal{F}(\xi)^{2} - 24\, \mathcal{F}(\xi)^{3} \right]=0;
\end{equation}
and a general solution for its linear part becomes
\begin{equation}\label{eq39c}
\mathcal{F}_0(\xi) =\tilde{C}_5\ {\mbox{$_2$F$_1$}\left(\frac{1}{2} \Delta_-,\frac{1}{4}-\frac{\nu}{2}; 1-\nu ;-{\xi}^{2}\right)}\,{\xi}^{\Delta_-} + \tilde{C}_6\ {\mbox{$_2$F$_1$}\left(\frac{1}{2} \Delta_+,\frac{1}{4}+\frac{\nu}{2}; 1+\nu ; -{\xi}^{2}\right)}\,{\xi}^{\Delta_+},
\end{equation}
where ${\displaystyle {}_{2}F_{1}(a,b;c;Z)}$ are the (Gauss's) \emph{hypergeometric functions}. Next, from the leading-order (LO) solution, one can get the higher order ones in generalized hypergeometric functions; However, for the specified bulk modes, after substitution $\xi = {u}/{r}$ and series expansion about $u=0$, we get
\begin{equation}\label{eq41f}
  f^{(1)}(u,r) = \left[ \hat{C}_{\Delta_+} + \check{C}_{\Delta_+}\,\ln(\frac{r}{u}) \right] \left( \frac{u}{r}\right)^{\Delta_+} \approx \hat{\beta}\, u^{\Delta_+} ,
\end{equation}
where we just kept the proper normalizable term for the corresponding boundary analyzes, with the interconnected real constants of $\hat{C}_{\Delta_+}$ and $\check{C}_{\Delta_+}$.

\subsection{General Procedure to Solve Equations with ADM}\label{sub02-02aa}
In the following three subsections, we use \emph{Adomian Decomposition Method}  \cite{Adomian1994} as a semi-analytical method to solve in particular NLPDE's; see also \cite{Wazwaz2009}. As the aim here is to have an expansion or a solution near the boundary ($u=0$) and the equation (\ref{eq02}) or (\ref{eq26}) is of the second order, we write $f_0(0,r)= f(0,r)+ u\, f_t (0,r)$ and can in general use one of the following initial data
\begin{equation}\label{eq40}
f(0,r)= {f}_-(r)\, u^{\Delta_-}, \quad  f(0,r)= {f}_+(r)\, u^{\Delta_+}, \quad f(0,r)= {f}_-(r)\, u^{\Delta_-} + {f}_+(r)\, u^{\Delta_+},
\end{equation}
corresponding to Neumann, Dirichlet and mixed BCs (that we use to link the bulk solutions to the boundary ones) from the left, respectively. It should also be noted that although arriving at a general solution valid for all modes with this method is rather difficult, nevertheless, as we will see, we can get interesting series solutions for specified bulk modes.

On the other hand, the equations we meet here comes from (\ref{eq02}) with or without $\delta$ and $F$ terms; and that for the common linear part, that is
\begin{equation}\label{eq50cc}
     \Box_4 f_0 - m^2\, f_0 =0,
\end{equation}
there is a closed solution in coordinate space as \cite{Witten}
\begin{equation}\label{eq50dd}
     f_0(u,\vec{u})= \bar{C}_{\Delta_+} \left(\frac{u}{u^2+(\vec{u}-\vec{u}_0)^2} \right)^{\Delta_+}, \quad \bar{C}_{\Delta_+}=\frac{\Gamma(\Delta_+)}{\pi^{3/2}\, \Gamma(\nu)}.
\end{equation}
Next, we take next to the boundary ($u=0$) behavior of the latter solution (\ref{eq50dd}), which according to (\ref{eq50ee}) has the structure of the middle expression in (\ref{eq40}), as the initial data or zeroth-order solution. Then, to perform the iteration process in the ADM and get perturbative solutions, we use
\begin{equation}\label{eq50ff}
\Box_4 f_{n+1} - m^2\, f_{n+1} = \sum_{n=0}^\infty A_n.
\end{equation}
where $A_n$'s are the Adomian polynomials \cite{Adomian1994}
\begin{equation}\label{eq50gg}
     A_n=\frac{1}{n!} \frac{d^n}{d\lambda^n} \left[\bar{F}\left(\sum_{l=0}^n \lambda^n\, f_n \right) \right]_{\lambda=0}, \quad n=0,1,2,... \ ,
\end{equation}
in which $\bar{F}(f)$ is for the nonlinear terms and so,
\begin{equation}\label{eq50hh}
     A_0=\bar{F}(f_0), \quad A_1= f_1\, \bar{F}^\prime (f_0), \quad A_2= f_2\, \bar{F}^\prime (f_0) + \frac{1}{2!} f_1^2\, {\bar{F}}^{\prime\prime} (f_0),... \ ,
\end{equation}
where primes mark differentiations wrt the argument, and the series expansions are written according to (\ref{eq28a}). Therefore, the Adomian polynomials for (\ref{eq02}), discarding the nonhomogeneous term $F$ for now because it just adds a constant non-dynamical term to the final solution, without and with the $\delta$ term, read
\begin{equation}\label{eq50jj}
     A_0=\lambda\, f_0^3, \quad A_1= 3\, \lambda\, f_0^2\, f_1, \quad A_2= 3\, \lambda\, f_0^2\, f_2 + 3\, \lambda\,f_0\, f_1^2, ... \ ,
\end{equation}
\begin{equation}\label{eq50kk}
    \begin{split}
    & \ A_0=\lambda\, f_0^3 - \delta\, f_0^2, \quad A_1= 3\, \lambda\, f_0^2\, f_1- 2\, \delta\, f_0^2\, f_1, \\
    & \quad A_2= 3\, \lambda\, f_0^2\, f_2 + 3\, \lambda\,f_0\, f_1^2 - 2\, \delta\, f_0\, f_2- \delta\, f_1^2 , ... \ ,
     \end{split}
\end{equation}
respectively.

\subsection{Solutions For the Massive \pmb{$m^2=4$} Equation with ADM}\label{sub02-02bb}
The massive $m^2=4$ mode could be realized with $C_1=1$ and $C_2=0$ in the WR version of (\ref{eq02}) as
\begin{equation}\label{eq08dd}
    \Box_4\, f - 4\, f - 24\, f^3 = 0,
\end{equation}
for which we have already presented rough solutions in \cite{Me6} and \cite{Me7}. In particular, with $C_1=1$ and $C_2=\sqrt{3}/12$ in the SW version of (\ref{eq02}) or (\ref{eq09}), we have
\begin{equation}\label{eq08ee}
    \Box_4\, f - 4\, f + 12\sqrt{3}\, f^2 - 24\, f^3 = 0,
\end{equation}
reminding that $F=0$.

Now, we use the initial data or near the boundary solution of
\begin{equation}\label{eq52aa}
    f_0(u\rightarrow0,r)= f(r)\, u^{4},
\end{equation}
which is indeed the middle condition of (\ref{eq40}) with $f_+(r) \equiv f(r)$ and $\Delta_+=4$, to get perturbative solutions for the equations (\ref{eq08dd}) and (\ref{eq08ee}) with ADM and the polynomials of (\ref{eq50jj}) and (\ref{eq50kk}), respectively. As a result, the series solutions about $u=0$, up to the second or next-to-next-to-leading order (NNLO), read
\begin{equation}\label{eq52bb}
   f^{(2)}(u,r)= f(r) \left[1- 6\,\ln(u) + 36\, \ln(u)^2 \right] {u}^{4} +  O({u}^{6}),
\end{equation}
\begin{equation}\label{eq52cc}
   f^{(2)}(u,r)= 3\,f(r)\, {u}^{4} -\frac{23}{200} \left(\frac{d^2 f(r)}{d r^2} + \frac{2}{r} \frac{d f(r)}{d r} \right) {u}^{6} + O({u}^{8}),
\end{equation}
respectively. Also, one may use near the boundary behavior of the solution of (\ref{eq50dd}), which is $f(r)={\bar{C}_4}/{r^8}$ for this case, as the initial data to rewrite the latter solutions hereon.

\subsection{Solutions For the Massless \pmb{$m^2=0$} Equation with ADM}\label{sub02-02}
The massless $m^2=0$ mode \footnote{It is notable that if we look at the mass of (\ref{eq08b}), we see that the first term on its RHS behaves like $\xi R_4$ with the conformal coupling $\xi=1/6$ and the Ricci scalar $\mathcal{R}_4=-12$ for $EAdS_4$, and the second term could be considered as $m_0^2$ of $ m_0^2 + \xi R_4 \equiv m^2$, which must in turn be 2 if $m^2=0$. However, according to the original literature on SUGRA, the 70 (pseudo)scalars of $\mathcal{N}=8$ multiplet from compactification of 11D SUGRA on $S^7$ (look for instance at \cite{Duff1986} as an original review referencing related studies, and \cite{Bianchi2} for a newer look at the spectrum) are massless in that $m_0^2=0$ and so, given the conformal coupling, one has the well-known CC (pseudo)scalar of $m^2=-2$.} is realized, for the SW version, with $C_1=1/3$ in (\ref{eq08b}) (in addition to $C_2=0$ in (\ref{eq02a})) and so,
\begin{equation}\label{eq08}
    \Box_4\, f - 24\, f^3 = 0, 
\end{equation}
which could also be read from (\ref{eq25bb}) with $m^2=\bar{m}^2=0$ and $\delta=0$. Next, a closed solution for its linear part, which we already faced in \cite{Me9} when taking the external-space backreaction, and also from (\ref{eq50dd}) with $\Delta_+=3$, reads
\begin{equation}\label{eq08f}
 {f}_0(u,\vec{u}) = \tilde{C}_{7} + \frac{\bar{C}_{3}\, u^3}{\left[u^2 + (\vec{u}-\vec{u}_0)^2 \right]^3}.
\end{equation}

Next, we use the ADM with (\ref{eq08f}) (setting $|\vec{u}-\vec{u}_0|=r$ and $\tilde{C}_{7}=0$ for simplicity) as the initial data and the Adomian polynomials from (\ref{eq50jj}), to find a solution for (\ref{eq08}). As a result, a solution after series expansion around $u=0$ up to the first iteration of the method, wrt (\ref{eq28a}), reads
\begin{equation}\label{eq52a}
     f^{(1)}(u,r) \approx \bar{C}_{3} \left[1 - 2\, \ln(u) \right] \left(\frac{u}{r^2} \right)^3.
\end{equation}

Still, as another way to realize the massless mode, one may consider the SW version of the original equation (\ref{eq02}) with $C_1=1$ and $C_2=1/12$, and so
 \begin{equation}\label{eq08aaa}
    \Box_4\, f + 12\, f^2 - 24\, f^3 = {2}/{9};
\end{equation}
Then, with the Adomian polynomials from (\ref{eq50kk}) (with $\lambda=24$, $\delta=12$) and ignoring the nonhomogeneous term $F=-2/9$, the NLO series solution reads  
\begin{equation}\label{eq52b}
     f^{(1)}(u,r) \approx \tilde{C}_{7} \left[1 - 6\, \tilde{C}_{7} + 12\, \tilde{C}_{7}^2 - 2\, \ln(u) \right] + \bar{C}_3 \left[1 - 12\, \tilde{C}_{7} + 36\, \tilde{C}_{7}^2 - 2\, \ln(u) \right] \left(\frac{u}{r^2} \right)^3.
\end{equation}

\subsection{Solutions For the Tachyonic \pmb{$m^2=-9/4$} Equation with ADM}\label{sub02-03a}
The so-called Breitenlohner–Freedman (BF) \cite{BreitenlohnerFreedman} bound of $m_{BF}^2=-9/4$ is the lowest physically acceptable bound for the (pseudo)scalar's mass in $AdS_4$ and so, it is interesting. It could be realized with $C_1=\frac{13}{12}$ \footnote{In fact, if we consider $(1 - 3\, C_1) \equiv \xi\, \mathcal{R}_4$, then according to the arguments in \cite{Hrycyna2017} for the nonminimal coupling value of $\xi=3/16$, $C_1={13}/{12}$ is realized in $EAdS_4$.} and $C_2=0$ in the SW version of (\ref{eq02}) as
\begin{equation}\label{eq08bb}
    \Box_4\, f + \frac{9}{4}\, f - 24\, f^3 = 0;
\end{equation}
Or with $C_2\neq 0$ in the SW background, for example with $C_1=2$ and $C_2= \sqrt{11}/(24 \sqrt{2})$ as
\begin{equation}\label{eq08cc}
    \Box_4\, f + \frac{9}{4}\, f + 3\sqrt{22}\, f^2 - 24\, f^3 = (49\sqrt22)/288;
\end{equation}
and we discuss their solutions in this subsection.

For the latter equations (\ref{eq08bb}) and (\ref{eq08cc}), we can use the initial data of
\begin{equation}\label{eq53aa}
    f_0(u\rightarrow0,r)= f(r)\, u^{3/2},
\end{equation}  
which is the middle condition of (\ref{eq40}) with $\Delta_+=3/2$, in ADM; In addition that we can read a similar initial condition, from the closed solution of (\ref{eq50dd}) for the linear parts of the equations, as
\begin{equation}\label{eq53bb}
    f_0(u,r)= \bar{C}_{3/2}\, \left( \frac{u}{u^2+r^2} \right)^{3/2} \Rightarrow f_0(u\rightarrow0,r) \approx \bar{C}_{3/2}\, \frac{u^{3/2}}{r^3}.
\end{equation}

Then, for the equation (\ref{eq08bb}), using (\ref{eq25bb}) and the Adomian polynomials from (\ref{eq50jj}), wrt (\ref{eq28a}), we get the third or NNNLO series solution of
\begin{equation}\label{eq53ff}
   f^{(3)}(u,r)= \bar{C}_{3/2} \left[ 1+\frac{\ln(u)}{4}+\frac{\ln(u)^2}{16}+\frac{\ln(u)^3}{64}\right] \left( \frac{u}{r^2} \right)^{3/2} + O({u}^{7/2}),
\end{equation}
with the initial data of (\ref{eq53bb}); and note that the higher degrees of logarithm appear for the higher orders of the expansion.

For the other equation (\ref{eq08cc}), we rewrite the homogeneous part of (\ref{eq25bb}) as
\begin{equation}\label{eq53hh}
        \begin{split}
        & \left(\partial_i \partial_i + \partial_u \partial_u \right)\, g_0 -\frac{(2+m^2)}{u^2}\, g_0 =0, \\
        & \left(\partial_i \partial_i + \partial_u \partial_u \right)\, g_{i+1} -\frac{(2+m^2)}{u^2}\, g_{i+1} =\sum_{i=0} A_i, \end{split}
\end{equation}
with the Adomian polynomials of
\begin{equation}\label{eq53ii}
     A_0 = -\frac{\delta}{u} g_0^2 + 24\, g_0^3 , \quad  A_1= -\frac{2 \delta}{u} g_0\, g_1 + 72\, g_0^2\, g_1, \ ... \, ,
\end{equation}
and $\delta = 3 \sqrt{22}$. Then, if we use $g_0= f(r)\, u^{1/2}$ from (\ref{eq53aa}) as the initial data or LO solution, we arrive at the first or NLO series solution of
\begin{equation}\label{eq53jj}
   f^{(1)}(u,r)= 3\,f(r)\, {u}^{3/2} -\frac{69\, \sqrt{22}}{16}\, f(r)^2\, u^3 - \frac{116}{225} \left(\frac{d^2 f(r)}{d r^2} + \frac{2}{r} \frac{d f(r)}{d r} \right) {u}^{7/2} + O({u}^{9/2}),
\end{equation}
near the boundary; and that, as an example, one may replace $f(r)={\bar{C}_{3/2}}/{r^3}$, as the initial data from (\ref{eq53bb}), in (\ref{eq53jj}).

\section{Likening the Bulk AdS$_4$ to the Boundary CFT$_3$} \label{sec04}
With respect to the ansatz (\ref{eq01}) and SW M2-branes background, the bulk (pseudo)scalars could be considered as arisen from membranes wrapping around mixed directions of the internal space $CP^3 \ltimes S^1/Z_k$, so that the resulting theory is for anti-M2-branes with braking all supersymmetries and parity invariance; and at the same time, the solutions of the bulk equation (\ref{eq02}) also break scale invariance and so, the conformal group $SO(4,1)$ of $EAdS_4$ breaks into $SO(4)$ (or $SO(3,1)$ of $dS_3$ with Lorentzian signature); see also \cite{Me7}. As a result and to find dual boundary solutions, we swap the three reps $\textbf{8}_s = \textbf{1}_{-2} \oplus \textbf{1}_{2} \oplus \textbf{6}_{0}$, $\textbf{8}_c = \textbf{4}_{-1} \oplus \bar{\textbf{4}}_{1}$, $\textbf{8}_v = \bar{\textbf{4}}_{-1} \oplus \textbf{4}_{1}$ of $SO(8) \rightarrow SU(4) \times U(1)$ so that, we can realize the needed singlet (pseudo)scalars and operators from $\textbf{35}_{s}\rightarrow \textbf{1}_{-4} \oplus \textbf{1}_{0} \oplus \bar{\textbf{1}}_{4} \oplus \textbf{6}_{-2}$ after the swapping and gauge field from $\textbf{28} \rightarrow \textbf{6}_{-2} \oplus \textbf{1}_{0} \oplus \bar{\textbf{6}}_{2} \oplus \textbf{15}_{0}$; see also \cite{Me9}. In addition, while we know that adding CS terms to the boundary $O(N)$ or $U(N)$ field theories can also account for parity breaking, such a breaking even causes the ABJM quiver gauge group of $SU(N)_k \times SU(N)_{-k}$ cuts down to one part, also because of the so-called novel Higgs mechanism \cite{ChuNastaseNilssonPapageorgakis}, as well as from the idea of fractional (anti)M2-branes \cite{ABJ} (as M5-branes wrapped around three internal directions; see the ansatz \ref{eq01}) so that with $l$ fractional (anti)M2-branes, the original gauge group is multiplied by $SU(l)$, which will in turn be in action while the original one remains as a spectator; see also \cite{Me5}.

On the other hand, we remind that a bulk (pseudo)scalar with near the boundary behavior of (\ref{eq50ee}) could be quantized with either Neumann ($\delta\beta =0$), Dirichlet ($\delta\alpha=0$) or mixed boundary condition; see, for instance, \cite{Balasubramanian} and \cite{KlebanovWitten}. In fact, while the regularity (of $\Delta_+$) and stability need the (pseudo)scalar mass is above the BF bound of $m_{BF}^2=-9/4$ in $AdS_4$, the Dirichlet or standard BC can be used for any mass; and at the same time, the Neumann or alternate BC is used for the masses in the range of $-9/4\leq m^2 \leq -5/4$ warranting stability as well \cite{BreitenlohnerFreedman02}. Meanwhile, settling $\alpha$ and $\beta$ as, respectively, source and vacuum expectation value for the one-point function of $\Delta_+$ operator and conversely for $\Delta_-$ operator, we notice that for $m^2 \geq -5/4$, just the $\beta$ mode is normalizable and the standard BC may be used, while for the negative range of the mass-squared, both modes are normalizable and all BCs may be used.

However, although in this study we mainly work with normalizable modes and the standard BC, the Euclidean AdS/CFT dictionary for both standard and alternate BCs reads
\begin{equation}\label{eq116}
   \begin{split}
   & \sigma \equiv \langle \mathcal{{O}}_{\Delta_+} \rangle_{\alpha} = - \frac{\delta W[\alpha]}{\delta\alpha} = \frac{1}{3} \beta, \quad W[\alpha] = -S_{on}[\alpha], \\
   & \ \ \ \ \ \ \langle \mathcal{{O}}_{\Delta_-} \rangle_{\beta} = - \frac{\delta \tilde{W}[\sigma]}{\delta\sigma}= \alpha, \quad \ \ \ \tilde{W}[\sigma] = - \tilde{S}_{on}[\sigma],  \\
   & \ \ \ \ \ \ \ \ \ \ \ \ \ \ \ \ \tilde{W}[\sigma] = - W[\alpha] - \int d^3 \vec{u}\ \alpha(\vec{u})\, \sigma(\vec{u}),
   \end{split}
\end{equation}
where $S_{on}$ and $\tilde{S}_{on}$ are on-shell actions in the bulk $AdS_4$, and $W[\alpha]$ and $\tilde{W}[\sigma]$ (as Legendre transform of the former) are generating functionals of the connected correlators of $\mathcal{{O}}_{\Delta_+}$ and $\mathcal{{O}}_{\Delta_-}$ on the boundary CFT$_3$, respectively.

\section{Boundary CFT$_3$ Duals For the Bulk Solutions}\label{sec05}
To find dual boundary solutions, we always use the elements of the ABJM action- see \cite{Terashima} and \cite{Me3}, \cite{Me4}- while keeping just the $U(1) \times U(1)$ part of the gauge group, unless otherwise stated. In addition, to have the needed $SU(4) \times U(1)$-singlet Lagrangians after the swappings we discussed in the previous section \ref{sec04}, depending on the case we consider just one scalar (say $Y= \varphi = h(r)\,\emph{\textbf{I}}_N$, with $h(r)$ as a scalar profile) or one fermion (say $\psi_{\hat{a}}^a= \delta_{\hat{a}}^a\, \psi$) and so, the ABJM scalar and fermion potentials vanish. Therefore, we employ the following Lagrangian:
\begin{equation}\label{eq56}
  \mathcal{L}^{(i)} = \mathcal{L}_{CS}+ \hat{\mathcal{L}}_{CS} - \texttt{tr} \left(i \bar{\psi} \gamma^k D_k \psi \right) - \texttt{tr}\left(D_k Y^{\dagger} D^k Y \right) - \mathcal{W}^{(i)},
\end{equation}
where $D_k \Phi = \partial_k \Phi + i A_k \Phi - i \Phi \hat{A}_k$ with $\Phi$ for both $\psi$ and $Y$, $F_{ij}=\partial_i A_j - \partial_j A_i + i \left[A_i, A_j \right]$, the last term $\mathcal{W}^{(i)}$, whose integral is equal to $W$ in (\ref{eq116}), serves as deformation, and we use the CS Lagrangian
\begin{equation}\label{eq56a}
  \mathcal{L}_{CS}^+ = \frac{i k}{4\pi}\ \varepsilon^{ij k}\ \texttt{tr} \left(A_i^+ \partial_j A_k^+ + \frac{2i}{3} A_i^+ A_j^+ A_k^+ \right),
\end{equation}
noting that we define $A_i^{\pm} \equiv (A_i \pm \hat{A}_i)$ and that, the matter fields of ABJM are neutral wrt $A_i^+$ (diagonal $U(1)$) while $A_i^-$ acts as baryonic symmetry, and since our (pseudo)scalars are neutral, we set $A_i^-=0$.

\subsection{Dual Solutions For the Massive Bulk State}\label{sub05-01}
For the nonminimally coupled (pseudo)scalar $m^2=+4$, besides the $\Delta_+=4$ operators of $\mathcal{O}_{4}^{(a)} =\texttt{tr}(\psi \bar{\psi})\, \texttt{tr}(\varphi \bar{\varphi})^2$ and $\mathcal{O}_{4}^{(b)} = \texttt{tr}(\psi \bar{\psi})^2$ in \cite{Me6}- see also \cite{Me7} and \cite{Me9}- here we introduce another interesting one as
\begin{equation}\label{eq57a}
 \mathcal{O}_{4}^{(c)}=\texttt{tr}(\psi \bar{\psi})\, \varepsilon^{ij}\, F_{ij}^+,
\end{equation}
which indeed accounts for the deformation in (\ref{eq56}) wrt (\ref{eq116}). Then, discarding the scalar kinetic term and with just the CS term of (\ref{eq56a}), the resulting equations for $\bar{\psi}$ and $A_i^+$ read 
\begin{equation}\label{eq57b}
 i \gamma^k \partial_k \psi + \psi\, \varepsilon^{ij}F_{ij}^+=0,
\end{equation}
\begin{equation}\label{eq57bb}
\varepsilon^{k ij}\, F_{ij}^+ =0,
\end{equation}
respectively, noting that the deformation term of (\ref{eq57a}) does not contribute to the gauge equation; and that because of working with normalizable modes and Dirichlet BC, we set $\alpha=1$ in the equations for simplicity. As a result, a solution for the $\psi$ equation, also in \cite{Me3}, \cite{Me6} and \cite{Me9}, reads
\begin{equation}\label{eq57c}
    \psi = \pm \tilde{a} \frac{\left({a} + i (x - x_0)^k \gamma_k \right)}{\left[{a}^2 + (x - x_0)_k (x - x_0)^k \right]^{\varsigma}} \left(\begin{array}{c}   1  \\   0   \end{array}\right),
\end{equation} 
where we use the Euclidean gamma matrices $\gamma^k=(\sigma_2,\sigma_1,\sigma_3)$, and $a=0$, $\varsigma=3/2$, $\tilde{a} = \frac{{i}}{2} \sqrt[3]{\frac{4}{5}}$ for the case at hand. On the other hand, for the gauge field equation, we consider the ansatz- see \cite{Seiberg-Witten1999} for a similar one- of
\begin{equation}\label{eq57d}
      A_i^+ = \varepsilon_{ij}\, x^j A(r), 
\end{equation}
where $A(r)$ is a scalar function on the boundary; and then, a suitable solution ($\sim r^{-\Delta_+}$) for the gauge equation of (\ref{eq57bb}) is accessible only if we take $x=y=z=r/\sqrt{3}$. However, if we consider the ansatz of
\begin{equation}\label{eq57e}
      A_\mu^+ = \omega_{\mu \nu}\, x^\nu A(r), \quad \omega_{\mu \nu}=
  \left\{ \begin{split}
  & 1 \ \ \ : \ \nu>\mu, \\
  & 0 \ \ \ : \ \nu=\mu,\ \ \mu,\nu \neq i, j ,
  \end{split} \right.
\end{equation}
for the $U(1)$ gauge field, where $\mu,\nu$ are also for the three boundary indices, as a result
\begin{equation}\label{eq57f}
      \varepsilon^{ij} F_{ij}^+ = -12 A(r) - 4\, r \acute{A}(r);
\end{equation} 
Next, we can write
\begin{equation}\label{eq57g}
      A(r)= \frac{{a}_1 + 4 {a}_2 r}{4 r^4} \Rightarrow F^+ = \frac{{a}_1}{r^4},
\end{equation}
which vanishes as $r\rightarrow \infty$, noting that ${a}_1, {a}_2, ... $ are boundary constants. Then, as a primary test of the correspondence according to (\ref{eq116}), we see that
\begin{equation}\label{eq58}
     \langle \mathcal{O}_{4}^{(c)} \rangle_{\alpha}= \frac{16}{125} \frac{{a}_1}{r^8} \cong f(r),
\end{equation}
with $f(r)$ in (\ref{eq52cc}), where one can also adjust $\bar{C}_4 = {16 {a}_1}/{125}$.

Likewise, if we consider both CS terms in (\ref{eq56}) with the deformation (\ref{eq57a}), while discarding the scalar kinetic term, combining the equations of $A_i$ and $\hat{A}_i$ gives
\begin{equation}\label{eq59a}
     \frac{i k}{4\pi} \varepsilon^{k ij}\, F_{ij}^+ + 2\, \bar{\psi} \gamma^k \psi= 0, \quad F_{ij}^-=0;
\end{equation}
Next, with $A_i^-=0$, the $\bar{\psi}$ equation is again (\ref{eq57b}), whose combination with (\ref{eq59a}) renders
\begin{equation}\label{eq60}
 \gamma^k \partial_k \psi + \frac{8 \pi}{k} \texttt{tr}(\psi \bar{\psi})\, \gamma^3 \psi =0,
\end{equation}
in which taking the third component of the gamma matrices in the second term is for compatibility with the solution we employ for $\psi$; Indeed, we can read a solution for (\ref{eq60}) from (\ref{eq57c}) with $\varsigma=1$, $\tilde{a} = \sqrt{\frac{-3 i k {a}}{8 \pi}}$ and then, from (\ref{eq57b}), we get
\begin{equation}\label{eq60b}
      F_{ij}^+ = \frac{3}{2} \varepsilon_{ij} \frac{{a}}{\left[{a}^2 + (x - x_0)_k (x - x_0)^k \right]};
\end{equation}
As a result, we see that
\begin{equation}\label{eq61}
     \langle \mathcal{O}_{4}^{(c)} \rangle_{\hat{\alpha}} = \frac{9 k}{8 \pi} \left( \frac{{a}}{{a}^2+(\vec{u} - \vec{u}_0)^2} \right)^2,
\end{equation}
which corresponds to the bulk near the boundary solution of (\ref{eq41f}) for $\Delta_+=4$ with $\check{C}_{4}=0$, $\hat{C}_{4}={9 k}/{8 \pi}$, of course in the limit of ${a}\rightarrow 0 , r\rightarrow \infty$, noting that we could also take $r=|\vec{u}-\vec{u}_0|$ (or $|x-x_0|$) with $\vec{u}_0$ (or $x_0$) as an arbitrary origin. \\
Besides, we compute the action value for the modification term of (\ref{eq57a}) based on the fermion solution from (\ref{eq57c}) and (\ref{eq60b}) as follows
\begin{equation}\label{eq62}
   {S}_{(4)}^{modi.} = - \frac{9 k\, {a}^2}{2} \int_0^\infty \frac{r^2}{\left( {a}^2+r^2 \right)^2}\, dr = - \frac{9 \pi}{8}{a},
\end{equation}
which is finite, showing an instanton with size ${a} \geq 0$ sitting at the origin ($\vec{u}_0=0$) of a three-sphere with radius $r$ at infinity ($S_\infty^3$).

\subsection{Dual Solutions For the Massless Bulk State}\label{sub05-02}
We have already studied the $\Delta_+=3$ operators of $\mathcal{O}_{3}^{(a)} =\texttt{tr}(\varphi \bar{\varphi})^3$, $\mathcal{O}_{3}^{(b)} =\texttt{tr}(\varphi \bar{\varphi}) \texttt{tr}(\psi \bar{\psi})$ and $\mathcal{O}_{3}^{(c)}=\texttt{tr}({A} \wedge {F})$, corresponding to the bulk massless $m^2=0$ mode, in \cite{Me3}, \cite{Me4}, \cite{Me5}, \cite{Me7} and \cite{Me8}. Here we consider another one as
\begin{equation}\label{eq63a}
 \mathcal{O}_3^{(d)}=\texttt{tr}(\varphi \bar{\varphi})\, \varepsilon^{ij}\, F_{ij}^+,
\end{equation}
and go through a similar procedure as for (\ref{eq57a}). In fact, discarding the fermion kinetic term and employing either the CS term of (\ref{eq56a}) or both of (\ref{eq56}), with $Y=Y^{\dag}$ and $A_i^-=0$, the $\bar{\varphi}=\varphi^\dag$ equation reads
\begin{equation}\label{eq63b}
 \partial_k \partial^k \varphi - \varphi\, \varepsilon^{ij}\, F_{ij}^+=0, 
\end{equation} 
and the gauge equation is the same as (\ref{eq57bb}) for which we take the solution of (\ref{eq57g}), while for the scalar part, we write
\begin{equation}\label{eq63c}
 \partial_k \partial^k h(r)=0 \Rightarrow h(r)= {a}_3 + \frac{{a}_4}{r};
\end{equation}
and see that with ${a}_3=0$, $\langle \mathcal{O}_{3}^{(d)} \rangle_{\alpha}\sim {1}/{r^6}$, which could in turn be adjusted with the bulk solutions of (\ref{eq52a}) and (\ref{eq52b}).

However, when considering both CS terms of (\ref{eq56}) or when there is only one gauge field, say $A_i$, but $Y\neq Y^{\dag}$ that in turn is reasonable because of being in Euclidean space, the gauge equation reads
\begin{equation}\label{eq64}
 \frac{i k}{4\pi} \varepsilon^{k ij} F_{ij} + i \left[Y \left(D^k Y^{\dag}\right)- \left(D^k Y \right) Y^{\dag}\right]=0,
\end{equation}
instead of (\ref{eq57bb}), with replacing $F_{ij}$ instead of $F_{ij}^+$ in (\ref{eq63a}). Next, we note that with $Y = Y^{\dag}$, we recover (\ref{eq57bb}), where we could also take
\begin{equation}\label{eq64a}
 F^{+}(u, \vec{u}) = \left(\frac{{a}}{{a}^2+(\vec{u}-\vec{u}_0)^2}\right)^2 = - h(u, \vec{u})^4,
\end{equation} 
recalling that $F^+(r\rightarrow \infty) \rightarrow 0$, \footnote{We first remind that the gauge solution of (\ref{eq64a}), already used in \cite{Me9} too, is indeed for the $SU(2)$ instanton originally studied in \cite{Belavin}. Second, we recall interpreting the added (anti)M5-branes wrapped around three internal directions and three external ones as fractional (anti)M2-branes and so, here we can have two fractional (anti)M2-branes and the gauge group of $SU(2)$; see the discussion in section \ref{sec04}.} and then the equation from (\ref{eq63b}) and its solution read 
\begin{equation}\label{eq64b}
 \partial_k \partial^k h(u, \vec{u}) + h(u, \vec{u})^5 =0 \Rightarrow  h(u, \vec{u}) = \left(\frac{{a}\, \sqrt{3}}{{a}^2+(\vec{u}-\vec{u}_0)^2}\right)^{1/2},
\end{equation}
which was also studied as an explicit $SO(4)$-invariant solution in \cite{Me8}. Next, as a primary test of the correspondence, we see
\begin{equation}\label{eq64c}
     \langle \mathcal{O}_{3}^{(d)} \rangle_{\alpha} = \sqrt{3} \left(\frac{{a}}{{a}^2+(\vec{u}-\vec{u}_0)^2}\right)^{3}, 
\end{equation}
which, with the Dirichlet BC, is structurally compatible with the bulk near the boundary solutions of (\ref{eq52a}) and (\ref{eq52b}), again in the limit of ${a}\rightarrow 0 , r\rightarrow \infty$. In addition, it may be considered as an instanton sitting at the conformal point of $u ={a}$ to match exactly with the original solution of (\ref{eq08f}) with $\tilde{C}_7=0$ and $\bar{C}_3=\sqrt{3}$. Meantime, the action value for the modification term of (\ref{eq63a}), with the solutions of (\ref{eq64a}) and (\ref{eq64b}), becomes
\begin{equation}\label{eq65}
   {S}_{(3)}^{modi.} = -4 \sqrt{3} \pi\, {a}^3 \int_0^\infty \frac{r^2}{\left( {a}^2+r^2 \right)^3}\, dr = - \frac{\sqrt{3}}{4} \pi^2,
\end{equation}
which is finite and again shows a (small) instanton at the origin of $S_\infty^3$.

Nevertheless, if we consider $Y= h(r)\, I_N$ and $Y^{\dag} = {a}_5\, I_N$ in (\ref{eq64}), next, from the gauge (\ref{eq64}) and  scalar (\ref{eq63b}) equations, we get 
\begin{equation}\label{eq66}
 \partial_k \left(\varepsilon^{k ij} F_{ij} \right) -\frac{4 \pi {a}_5}{k}\, h(r)\, \varepsilon^{ij} F_{ij}=0;
\end{equation}
Then, with $A_i$ of (\ref{eq57d}) and so $\varepsilon^{ij} F_{ij}= -4 A(r)$, we can write 
\begin{equation}\label{eq66a}
 \frac{dA(r)}{dr}+ h(r)\, A(r)=0, \quad {a}_5 \equiv -\frac{3 \sqrt{3}\, k}{4 \pi},
\end{equation}
and with
\begin{equation}\label{eq66b}
 h(r)= \frac{n}{r} \Rightarrow A(r)=  \frac{{a}_6}{r^n},
\end{equation}
where $n$ is a real number. As a result, with $n=4$, we again have $\langle \mathcal{O}_{3}^{(d)} \rangle_{\alpha}\sim {1}/{r^6}$; and with $n=1$, we have $\langle \mathcal{O}_{3}^{(d)} \rangle_{\hat{\alpha}} \sim {1}/{r^3}$, which in turn corresponds to the bulk solution of (\ref{eq41f}) for $\Delta_+=3$ with $\check{C}_{3}=0$ and $\hat{C}_{3} = {a}_6$.

Likewise, with $A_{i}$ of (\ref{eq57e}) and then (\ref{eq57f}), from (\ref{eq66}) we can write
\begin{equation}\label{eq67a}
  \frac{d^2 A(r)}{dr^2} + \left( h(r)+ \frac{4}{r} \right) \frac{dA(r)}{dr}+ \frac{3}{r}\, h(r)\, A(r)=0,
\end{equation}
with the same ${a}_5$ as in (\ref{eq66a}) and next, with $h(r)$ of (\ref{eq66b}), we get
\begin{equation}\label{eq67b}
  A(r) = \frac{{a}_6}{r^n} + \frac{{a}_7}{r^3} \Rightarrow \varepsilon^{ij}F_{ij} \equiv F =\frac{4\, {a}_6}{r^n} (n-3);
\end{equation}
and then, with $n=4, 1$, we have the same correspondences stated below the solution of (\ref{eq66b}).

\subsection{Dual Solutions For the Tachyonic Bulk State}\label{sub05-03}
We can build the wished $\Delta_+=3/2$ operators, corresponding to the bulk BF $m^2=-9/4$ mode, out of the scalar, fermion and gauge field we use in this section. For instance, one may consider
\begin{equation}\label{eq68a}
     \mathcal{O}_{3/2}^{(a-d)} = \texttt{tr}(\varphi {\psi}), \texttt{tr}(\bar{\varphi} \bar{\psi}), \texttt{tr}(\bar{\varphi} {\psi}), \texttt{tr}({\varphi} \bar{\psi}),
\end{equation}
which have recently used in \cite{Aharony2011}, \cite{Chang2013}, \cite{Hikida-Wada2017} and particularly in \cite{Giombi2016} to set up duals to type-AB Vasiliev’s HS theories. In particular, we may employ the already mentioned operator $\mathcal{O}_{3}^{(b)}$ in subsection \ref{sub05-02} to build a desired dimension-3/2 operator as
\begin{equation}\label{eq68b}
     \mathcal{O}_{3/2}^{(e)} = \left( \texttt{tr}(\varphi \bar{\varphi}) \texttt{tr}(\psi \bar{\psi}) \right)^{1/2}.
\end{equation}
Having the latter operators of (\ref{eq68a}) and (\ref{eq68b}) to deform (\ref{eq56}) with, and the CS term of (\ref{eq56a}), $A_i^-=0$ and $\varphi=\bar{\varphi}$, the resulting equations satisfy if
\begin{equation}\label{eq69}
    \partial_k \partial^k h =0, \qquad i \gamma^k \partial_k \psi =0,
\end{equation}
with the solutions of (\ref{eq63c}) and (\ref{eq57c}) for the scalar and fermion, respectively. As a result, with near the boundary behavior of a bulk solution as $f(u\rightarrow 0) \approx u^{3/2} [\alpha \ln(u) + \beta]$, for the normalizable mode and Dirichlet BC, we have
\begin{equation}\label{eq70}
     \langle \mathcal{O}_{3/2}^{(e)} \rangle_{\alpha}= \frac{1}{2} \sqrt[3]{\frac{4}{5}} \frac{{a}_4}{r^3} \cong f(r),
\end{equation}
with ${a}_3=0$ of (\ref{eq63c}) and $f(r)$ of (\ref{eq53jj}); and in particular with ${a}_4 \cong 2\, \bar{C}_{3/2}$ for the bulk near the boundary solution of (\ref{eq53ff}). Further, we remind when considering both CS terms in (\ref{eq56}) and $A_i^-=0$, the $A_i^+$ equation (\ref{eq59a}) with the solution of (\ref{eq57c}), result in a zero magnetic charge or flux: $\Phi=\oint_{S^3_\infty} F^+=0$ \cite{Me3}.

As another deformation, we consider
\begin{equation}\label{eq71a}
 \mathcal{O}_{3/2}^{(f)}= \varepsilon^{k ij} \varepsilon_{ij}\, A_k^+\, \bar{\varphi};
\end{equation}
and then, without the fermion in (\ref{eq56}) and with just $\mathcal{L}_{CS}^+$ of (\ref{eq56a}), the equations for $\bar{\varphi}$ and $A_k^+$ read
\begin{equation}\label{eq71b}
 \partial_k \partial^k \varphi -\varepsilon^{k ij} \varepsilon_{ij}\, A_k^+=0, \quad \frac{i k}{4\pi}\, \varepsilon^{k ij} F_{ij}^+ -\varepsilon^{k ij} \varepsilon_{ij}\, \bar{\varphi}=0,
\end{equation}
from the left respectively; and from their combination we get
\begin{equation}\label{eq71c}
 \bar{\varphi}\, \partial_k \partial^k \varphi - \frac{i k}{4\pi}\, \varepsilon^{k ij}  A_k^+\, F_{ij}^+ =0,
\end{equation}
whose scalar part can be satisfied with (\ref{eq63c}) and its gauge part with a solution like 
\begin{equation}\label{eq71d}
 A_k^+ = \varepsilon_{k ij}\, \varepsilon^{ij} A^+(r),
\end{equation}
where $A^+(r)$ is another boundary scalar function. In particular, with $A^+(r) \sim 1/r^2$ and $h(r) \sim 1/r$, the basic correspondence $\langle \mathcal{O}_{3/2}^{(f)} \rangle_{\alpha} \sim 1/r^3$ with the bulk solutions of (\ref{eq53ff}) and (\ref{eq53jj}) (with $f(r)\sim 1/r^3$) is realized. Meanwhile, the same correspondence can be realized for the deformation operators of
\begin{equation}\label{eq72}
 \mathcal{O}_{3/2}^{(g)}= \texttt{tr}(\varphi \bar{\varphi})^{1/2} \varepsilon^{k ij} \varepsilon_{ij}\, A_k^+, \quad \mathcal{O}_{3/2}^{(h)}= \left( \texttt{tr}(\varphi \bar{\varphi})\, \varepsilon^{ij} F_{ij}^+ \right)^{1/2},
\end{equation}
where the same solutions as for (\ref{eq71a}) could be used for the first one on the left, while for another one, one may employ the scalar and gauge solutions of \ref{eq63c}) and (\ref{eq57g}) respectively.

Further, as the operators composed of the fermion and gauge fields, we consider
\begin{equation}\label{eq73}
 \mathcal{O}_{3/2}^{(p)}= \left( \texttt{tr}(\psi \bar{\psi})\, \varepsilon^{k ij} \varepsilon_{ij}\, A_k^+ \right)^{1/2}, \quad \mathcal{O}_{3/2}^{(q)}= \texttt{tr}(\psi \bar{\psi})^{1/4}\, \varepsilon^{k ij} \varepsilon_{ij}\, A_k^+,
\end{equation}
for the deformation in (\ref{eq56}) without scalars. Then, after some mathematical manipulations on the resultant fermion and gauge equations, we arrive at
\begin{equation}\label{eq73a}
 ({n}) i \bar{\psi} \gamma^k \partial_k \psi + \frac{i k}{4\pi}\, \varepsilon^{k ij} A_k^+ F_{ij}^+ + 2\, \bar{\psi} \gamma^k \psi\, A_k^+ =0,
\end{equation}
where ${n}=1, 4$ for $\mathcal{O}_{3/2}^{(p)}, \mathcal{O}_{3/2}^{(q)}$ respectively, and the last term exists when taking both CS terms in (\ref{eq56}) instead of just $\mathcal{L}_{CS}^+ $ of (\ref{eq56a}) and of course with $A_i^-=0$ again. In fact, for the latter case, without the last term of (\ref{eq73a}), the equation is satisfied with (\ref{eq71d}) and (\ref{eq57c}) and so, with $A^+(r) \sim 1/r^2$, we again have the basic correspondence of $\langle \mathcal{O}_{3/2}^{(p), (q)} \rangle_{\alpha} \sim 1/r^3$ wrt the bulk near the boundary solution of (\ref{eq53bb}). However, for the whole equation (\ref{eq73a}), given the ansatz of (\ref{eq71d}) for the gauge field, the equation is satisfied with $\psi$ from (\ref{eq57c}) with $\varsigma=3/2$ and $A^+(r)$ from (\ref{eq60b}) with $F_{ij}^+ \rightarrow A^+(r)$, $\varepsilon_{ij} \rightarrow \bar{n}$; and then
\begin{equation}\label{eq73c}
     \langle \mathcal{O}_{3/2}^{(p), (q)} \rangle_{\alpha} = \frac{{a}_8}{\left[ {a}^2+(\vec{u}-\vec{u}_0)^2 \right]^{3/2}},
\end{equation}
with $\bar{n}=1, 2$  and ${a}_8 = (\sqrt{18 {a}}) \tilde{a}, (18 \bar{n} {a}) \sqrt{\tilde{a}}$ for $\mathcal{O}_{3/2}^{(p)}, \mathcal{O}_{3/2}^{(q)}$ respectively. As a result, we again see the operator-state correspondence with the bulk near the boundary solutions in subsection \ref{sub02-03a}, for the normalizable mode and Dirichlet BC, and in the limit of ${a}\rightarrow 0$ and $r\rightarrow \infty$. 

Besides, the finite value of the suiting action from (\ref{eq56}), without the scalar term, with the deformation of (\ref{eq73}) and based on the solutions from (\ref{eq57c}) and (\ref{eq60b}), noting that the contribution of the CS part vanishes, with the help of the equation (\ref{eq73a}), becomes
\begin{equation}\label{eq73d}
   {S}_{(3/2)}^{modi.}= \frac{24 \pi\, {a}\, \tilde{a}^2\, \bar{n}}{{n}} \int_0^\infty \frac{r^2}{\left( {a}^2+r^2 \right)^3}\, dr = \frac{3 \pi^2 \bar{n}}{2 {n}} \frac{\tilde{a}^2}{{a}^2},
\end{equation}
which, similar to (\ref{eq62}) and (\ref{eq65}), shows the (small) instanton nature of the solution.

\section{Conclusion and Comments}
From a reduction of 11D SUGRA over $AdS_4 \times CP^3 \ltimes S^1/Z_k$, we got a scalar NLPDE in the bulk of Euclidean $AdS_4$. As the main bulk equation, valid in probe approximation, we focused on three massive, massless and tachyonic (pseudo)scalar modes of $m^2=+4, 0, -9/4$ of it, and tried to present solutions suitable for near the boundary Euclidean CFT$_3$ analyzes. Meantime, for the anti-M2-branes background, the equation could be homogeneous with Higgs-like (pseudo)scalars that provided spontaneous symmetry breaking; and for that we presented some perturbative solutions. 

In particular, we employed the Adomian decomposition method as a semi-analytical method to solve nonlinear differential equations, with the initial data of near the boundary behavior of a general solution and a free massive one in the bulk. In that way, we got interesting perturbative series solutions (mainly normalizable with Dirichlet BC) near the boundary for the three bulk modes.

On the other hand, given the fact that the bulk setups and solutions break all supersymmetries, parity- and scale-invariance, and the resultant theory was for anti-M2-branes, we interchanged the three reps of $SO(8)$ for gravitino and built new $SU(4) \times U(1)$-single $\Delta_+=4,3,3/2$ operators out of a scalar, fermion and gauge fields of an ABJM-like 3D CS $SU(N)$ field theory. Next, by deforming the suiting boundary actions by the operators and solving the resulting equations, we found (small) instanton solutions on a three-sphere of the radius $r$ at infinity. Then, we confirmed the state-operator AdS$_4$/CFT$_3$ correspondence for all the bulk and boundary solutions we presented.

It should be noted that, as far as I know, the ADM \cite{Adomian1994} is not used so far to solve equations in the framework of  AdS/CFT correspondence and so, it will be interesting to explorer it more.

And as a final matter related to the studies here, it is also interesting to investigate hadron and specially meson spectra in framework of AdS$_4$/CFT$_3$ correspondence in addition to the roles of instantons. In this way, one usually embeds special branes in high-dimensional supergravity side of the model (through the Dirac-Born-Infeld action) and study (scalar) fluctuations around the probe branes.\footnote{It is noteworthy that in the commonly used bottom-up approach, including hard-wall \cite{Polchinski-Strassler2001} and soft-wall \cite{Karch2006} approximations, people often start from the gravity models in AdS and employs AdS/CFT dictionary to get the content of dual QCD; noting that in the soft-wall model, the conformal (scale) symmetry is broken by the bulk scalar dilaton (as it is in our top-down approach) and the scale $\Lambda_{QCD}$ is introduced-See also \cite{Colangelo2008}.} Then, after solving the Schrodinger-like equations from the corresponding bulk action, quark masses and condensates are derived from the asymptotic behavior of the solutions (e.g. hadrons correspond to normalizable solutions of the bulk equations; and generally, bulk scalar fields account for quark and gluon condensates). In other words, the vacuum expectation values of corresponding boundary operators and their correlations give masses, coupling constants and form factors of hadrons - See, for instance, \cite{Csaki1998}, \cite{Kruczenski2003}, \cite{Sakai-Sugimoto2004} among original studies and \cite{Erdmenger2007} for a review with references therein; And for brane embedding in ABJM and corresponding meson spectra, see for instance \cite{Hikida} and \cite{Bea2014}, where in the former a scalar meson operator like our $\mathcal{O}_{4}^{(a)}$ is introduced.\\
In particular, corresponding to probe branes embedded in gravity side, the so-called F-term $m_q \texttt{tr}(\psi \bar{\psi})$-look also at \cite{Me9} where such a term is present as a deformation in the context of regular and critical fermion model- breaks the chiral symmetry (into $SU(3)_V$), causes quarks to have mass and so, there will be eight pseudo-Goldstone bosons (considered as pseudo-scalar mesons). To see the corresponding operators of QCD$_4$, look for instance at \cite{Teramond-Brodsky2005}, \cite{Erlich2005}, \cite{Rold-Pomarol2005} and \cite{Jugeau2009}. Similarly, if our setup was right for QCD$_3$, the operators here might account for various (scalar) mesons and glueballs (quark and gluon condensates), regarding the roles of instantons as non-perturbative (gluon) effects in chiral symmetry breaking, strong interaction among quarks and hadron spectra- for a review on "instantons in QCD", see \cite{Schaefer-Shuryak1996}; For example, the operator $\mathcal{O}_4^{(c)}$ might be considered for quark-antiquark interaction through instantons/gluons. It will be interesting to go further in this direction.

\section{Acknowledgments}
I would like to thank the members (in particular A. Dabholkar and K. Papadodimas) of the High Energy, Cosmology and Astroparticle Physics (HECAP) section of the Abdus Salam International Centre for Theoretical Physics (ICTP), where a part of this work was done during my recent visit there, for scientific discussions and hospitality. I would also like to thank A. Imaanpur for pointing out a subtle point in a part of the calculations.



\end{document}